# A national-scale cross-time analysis of university research performance[1]

*Giovanni Abramo*[a,b,*], *Ciriaco Andrea D'Angelo*[a] *and Flavia Di Costa*[a]

[a] Laboratory for Studies of Research and Technology Transfer
School of Engineering, Dept of Management
University of Rome "Tor Vergata"

[b] National Research Council of Italy

**Abstract**

Research policies in the more developed nations are ever more oriented towards the introduction of productivity incentives and competition mechanisms intended to increase efficiency in research institutions. Assessments of the effects of these policy interventions on public research activity often neglect the normal, inherent variation in the performance of research institutions over time. In this work, we propose a cross-time bibliometric analysis of research performance by all Italian universities in two consecutive periods (2001-2003 and 2004-2008) not affected by national policy interventions. Findings show that productivity and impact increased at the level of individual scientists. At the level of university, significant variation in the rank was observed.

## 1. Introduction

With the aim of fostering greater efficiency in publicly funded research institutions, an increasing number of nations are introducing or strengthening their incentive systems and mechanisms for competition.

These include systems of selective funding for research that are based on results from national research assessment exercises, or on evaluation of project proposals. After first examining aspects of the methodology and the applications for analysis of productivity and impact of research activities in universities (Gómez et al., 2009; Abramo et al., 2008a; Bonaccorsi et al., 2006), scholars in the field have recently turned their attention to the evaluation of effects from the actual conduct of national research evaluation exercises.

Butler (2003) studied the effects of the criteria used to allocate resources to the Australian higher education system. In the 1990s, performance assessment exercises used publication counts as a key criterion. Significant funds were assigned on the basis of aggregate publication counts, with little attention to quality of output. There was, as a result, an increase in publication productivity between 1990 and 1998, but also a corresponding drop in relative quality at the international level. Moed (2008) examined the results of scientific activity in the UK, over the period of 1985 to 2004, in function of a sequence of various research assessment exercises (RAEs). The results show that research staff tend to orient their activity according to the guidelines of the evaluation exercises. In the 1992 RAE, the emphasis was on the quantitative aspect of scientific production, and examination showed a responding increase in the number of publications. However, when in 1996 the focus shifted from "output counts" to "quality", there was a greater tendency towards publication in journals with a higher impact factor. These two studies thus reveal a clear influence of the structure of the incentive system on researchers' behavior. Auranen and Nieminen (2010) arrive at different conclusions in their analysis of the variation in funding environments for university research across eight nations. They encounter significant differences in the competitiveness of funding systems but do not detect any "straightforward connection between financial incentives and the efficiency of university systems". In our judgment, their conclusions may be distorted by the absence of field standardization in the bibliometric elaboration of performance, as applied by the authors.

Studies evaluating the impact of assessment exercises, and selective research funding policies in general, present two quite significant problems. The first concerns the selection of the guidelines or "rules of the game" of the evaluation exercises, or of changes in the rules – events which are often not communicated to scientists until the research period that is the object of evaluation has already begun, or even not until finished. It is obviously not an easy task to achieve a distortion-free evaluation of effects from a research assessment exercise when such evaluation is based on results over a period of time when there are changes in methodologies, criteria or indicators of the exercises. The second problem is that of separating the effects of policy interventions, such as those correlated to results from assessment exercises, from endogenous factors, such as the normal, inherent variations of research institution performance over time. In fact, some of the changes observed in an organization can be independent of decision-making of both internal and external actors (Hung et al., 2009). The current study focuses on this second aspect, and we will provide a contribution on the first issue in a subsequent work.

To achieve the objective of measuring the natural variation in performance by research institutions, it is necessary that: i) for two successive periods under analysis, there are no intervening factors that could have altered resources dedicated to research, such as selective funding; and ii) the evaluation methodology and criteria are the same over both periods. The Italian case lends itself to this type of analysis, since, until 2008, government financing for universities was allocated solely on the basis of criteria for satisfying the resource needs of each university, and it was only beginning in 2009 that a part of state funds were assigned based on merit. In this work, we examine research performance in the hard sciences by Italian universities over two consecutive periods: 2001-2003 and 2004-2008. The choice of these intervals is motivated, in part, by the fact that a national evaluation exercise is about to be launched for the precise period of 2004-2008. The bibliometric performance analysis of these five years will doubtless also be useful in future to check for potential correspondence between the rankings generated by our analysis and those of the national evaluation exercise. The length of the first time period is determined in part by availability of data, which are not accessible for years prior to 2001. Performance will be measured for both periods by using the same bibliometric indicators i.e., publications and citation counts. Since neither time period involved policy interventions that altered the resources for research in the universities, the exercise will permit an understanding of the extent of natural variation in performance of the research organizations. Such variation should be subtracted from that detected subsequent to policy interventions, in order to evaluate the real impact of such interventions.

The next section of this work deals with the presentation of the dataset and methodology used. Section 3 presents and comments on the results obtained from the analyses conducted at the aggregate level, for the two time periods considered. Section 4 presents the results at the level of disciplines, while Section 5 details the analysis by single subfields falling under the same discipline. Section 6 presents a cross-section of the field-data, intended to show potential shifts in emphasis between quality and quantity. The final section offers a summary of the principal results and the authors' considerations.

**2. Data and method**

To reach the research objectives that inspire this work, it was necessary that the authors make precise methodological choices and take appropriate precautions to avoid risks connected to large scale research performance assessment (van Raan, 2005).

In the Italian university system, each researcher is an assigned member of a single scientific disciplinary sector (SDS). The SDSs are in turn grouped into 14 university disciplinary areas (UDAs). The field of observation for this study is composed of all Italian universities active in at least one of the 205 SDSs that make up the so-called "hard sciences", which compose 9 UDAs: mathematics and computer science, physics, chemistry, earth sciences, biology, medicine, agricultural and veterinary sciences, civil engineering and architecture, industrial and information engineering[2].

The dataset used for the analysis is based on *Observatory of Public Research in Italy* (ORP)[3], a database derived by the authors from the Thomson Reuters Web of Science™ (WoS). The ORP indexes all the scientific[4] and patent production by Italian public research organizations beginning from 2001. Taking the raw WoS data, and then applying complex algorithms for the reconciliation of the author's affiliation and the disambiguation of the true identity of the authors, each publication is attributed to the university scientists that produced it (Giuffrida et al., 2010). It is thus possible to associate the publications with the SDSs to which the authors belong. The SDSs are the unit of analysis for this study, however the data are ultimately presented at the level of the UDA. This is accomplished by standardizing and weighting the data referring to the SDSs that compose each UDA, so as to obtain robust ratings with respect to the intrinsic heterogeneity of the SDSs. In particular, field standardization is necessary for eliminating bias due to the different prolificacy levels (publication and citation intensity) of SDSs within a single UDA, while data weighting takes account of the differing representativity, in terms of members, of the SDSs present in each UDA (Abramo et al., 2008b).

As will be better shown below, several bibliometric indicators are used, referring to the number of publications, number of co-authors for each, and their impact in terms of citations received.

## 3. Cross-time analysis at the aggregate level

During the two periods, the research staff[5] of the 205 SDSs under observation increased considerably, going from 39,309 in 2001-2003 to 43,223 in the 2004-2008 period. The net increase of 3,914 is due to entry of 6,397 new researchers against 2,483 departures. Initially we will present an analysis at the national aggregate level of the variations of productivity and impact for the two periods. The results show an average increase in productivity of 20.6%: the average number of annual publications per researcher[6] is 1.513 for 2001-2003 and 1.825 for 2004-2008 (Table 1). The increase concerns all the UDAs: Agricultural and veterinary sciences shows the maximum increment (+45.8%), and Physics the minimum (+9.1%). The variation in Earth sciences is notable, where the average annual of output per researcher increases by a third.

The next question is whether this increase could be accompanied by a reduction of average impact in the scientific production of Italian researchers, as Butler (2003) detected, though under very different circumstances, in the Australian case.

To respond, calculations were completed for the average standardized impact[7] of the publications produced in the two periods. The data, grouped by the UDAs to which the

---

[2] In Italy there are 95 universities. Of these, only 63 (which are the units of analysis for this study) have at least 6 research staff in at least one of the UDA considered.
[3] www.orp.researchvalue.it
[4] Articles, reviews and conference proceedings.
[5] The research staff considered are assistant, associate and professors.
[6] This data was obtained as a weighted average of the average productivity of the 205 SDSs considered, with the weightings based on the portion of the national research staff belonging to each SDS.
[7] Here, the impact of a publication is standardized with respect to a WoS international benchmark: the

scientists belong, are presented in Table 2. In proceeding through time from the first to the second period, there is a substantial increase in average impact, both at the general level (+45.1%) and at the level of single UDA: Medicine shows the greatest increment (+62.4%), followed by Mathematics and computer science, (+56.7%) and Biology (+51.9%). The increase in average impact is never less than 28%. Finally, in the 2001-2003 triennium, there were 6 UDAs with an aggregate value of average impact that exceeded one, or the global average, while over the next five years this occurs in all the nine UDAs.

The variations observed can be traced to the variation in scientific performance of the personnel that remained active over both periods (93.7% of total) and the difference in performance between departing researchers and new entrants.

| UDA | 2001-2003 | 2004-2008 | Var. (%) |
|---|---|---|---|
| Mathematics and computer science | 0.932 | 1.166 | 25.1 |
| Physics | 2.689 | 2.935 | 9.1 |
| Chemistry | 2.768 | 3.115 | 12.5 |
| Earth sciences | 0.818 | 1.090 | 33.3 |
| Biology | 1.719 | 1.931 | 12.3 |
| Medicine | 1.559 | 1.975 | 26.7 |
| Agricultural and veterinary sciences | 0.846 | 1.256 | 48.5 |
| Civil engineering and architecture | 0.290 | 0.347 | 19.7 |
| Industrial and information engineering | 1.649 | 2.095 | 27.0 |
| Weighted average | 1.513 | 1.825 | 20.6 |

*Table 1: Output per researcher for the two periods under observation*

| UDA | 2001-2003 | 2004-2008 | Var. (%) |
|---|---|---|---|
| Mathematics and computer science | 1.044 | 1.636 | 56.7 |
| Physics | 1.153 | 1.482 | 28.5 |
| Chemistry | 1.127 | 1.669 | 48.1 |
| Earth sciences | 1.012 | 1.429 | 41.2 |
| Biology | 1.001 | 1.521 | 51.9 |
| Medicine | 1.021 | 1.658 | 62.4 |
| Agricultural and veterinary sciences | 0.914 | 1.213 | 32.7 |
| Civil engineering and architecture | 0.901 | 1.179 | 30.9 |
| Industrial and information engineering | 0.903 | 1.160 | 28.5 |
| Weighted average | 1.029 | 1.493 | 45.1 |

*Table 2: Aggregate values of average standardized impact per UDA for the periods 2001-2003 and 2004-2008*

## 4. Comparison between rankings at the UDA level

In the preceding section we showed that, between two successive time periods, there was a substantial increase in the bibliometric performance of the Italian academic system, both in terms of productivity and impact. Now we propose to evaluate the extent of the variations for the individual universities and, in particular, the shifts in ranking with reference to various bibliometric indicators. The performance of each researcher is measured using four indicators, calculated at the level of SDS, as follows:
- Productivity (P): total of publications authored by scientists of the SDS of the

average value of the citations of all publications indexed in the WoS in the same subject category and year.

university averaged over research staff of the SDS of that university;
- Fractional Productivity (FP): total of contributions to publications (averaged over research staff of the SDS of the university), with "contribution" depending on: i) the number of co-authors; and in the case of the life sciences, ii) each author's position in the listing; iii) the character of the co-authorship (intra-mural o extra-mural);
- Average Quality (AQ): the mean value of standardized citations to all publications authored by scientists of the SDS of the university. Citations of a publication are standardized dividing them by the median[8] of citations[9] of all Italian publications of the same year and WoS subject category;
- Fractional Scientific Strength (*FSS*): as for fractional productivity, but considering standardized citations to each publication authored by the SDS research staff.

As stated previously, the results of the analysis will be presented at the level of UDA. For this, the performance measures referring to SDS are first standardized in order to take into due account the different prolificacy levels (publication and citation intensity) of SDSs within a single UDA. The performance measures are then weighted to take account of the differing representativity, in terms of researcher numbers, of the SDSs belonging to each UDA.

For each of the two time periods considered (2001-2003) and (2004-2008), four ranking lists are calculated, one for each of the indicators. These ranking lists exclude universities from any particular UDA where, for that UDA, the university had a research staff of less than 6 individuals. For those remaining, the shifts in rank (2004-2008 vs. 2001-2003) were calculated as absolute values, for the four indicators. The statistics referring to the changes in ranking for P and FSS are presented in Table 3 and Table 4. The results for the other two indicators are similar. In general, almost all the universities vary in rank between the two periods under observation. Chemistry shows the lowest percentage of universities with a change of rank, being 84,4% of the total; Earth sciences shows the opposite case, with 100% of the 39 observed universities changing position in the rankings for the two periods.

| UDA | Variations | Max | Average | Median |
|---|---|---|---|---|
| Mathematics and computer science | 48 out of 51 (94.1%) | 41 | 9.41 | 8.0 |
| Physics | 44 out of 45 (97.8%) | 20 | 6.36 | 5.0 |
| Chemistry | 38 out of 45 (84.4%) | 27 | 6.71 | 5.0 |
| Earth sciences | 39 out of 39 (100.0%) | 30 | 7.92 | 6.0 |
| Biology | 47 out of 50 (94.0%) | 19 | 5.58 | 3.5 |
| Medicine | 40 out of 42 (95.2%) | 27 | 5.21 | 4.0 |
| Agricultural and veterinary sciences | 27 out of 29 (93.1%) | 16 | 5.83 | 6.0 |
| Civil engineering and architecture | 37 out of 38 (97.4%) | 19 | 6.74 | 5.5 |
| Industrial and information engineering | 44 out of 45 (97.8%) | 29 | 7.58 | 7.0 |

*Table 3: Statistics for the differences in ranking for P for universities between 2004-2008 and 2001-2003*

---

[8] The choice to standardize citations with respect to median value (rather than to the average, as is frequently observed in the literature) is justified by the fact that the distribution of citations is highly skewed in almost all subject categories (Lundberg, 2007). In the aggregate analysis presented in the previous section it was not possible to utilize the average, because the international average was not available, nor was it opportune to use the national average, since aggregate value would clearly result as invariant.
[9] Observed as of 30/06/2009.

Mathematics and computer science shows the greatest absolute value for change in rank: one university loses a full 41 positions. The agricultural and veterinary sciences UDA registers the lowest maximum for change in rank (16 positions). In the other areas, substantial changes are encountered in Earth sciences and Industrial and information engineering, with the maximum increment in rank being 30 and 29, for the two respective UDAs. For all the areas, with the exception of Agricultural and veterinary sciences, the average value of shift in rank is greater than the median, indicating that the distribution of the differences is asymmetric to right. Mathematics and computer science is the area with least alignment between the sets of rankings for 2001-2003 and 2004-2008, as seen both in terms of average change (9.41) and median change (8). The areas where the ranks experience the least variation are Medicine (average 5.21), Biology (5.58) and Agricultural and veterinary sciences (5.83).

For the FSS indicator, the percentages of total universities that show variations fall in a range between 88.9% for Physics and 98.0% for Mathematics and computer science. Mathematics and computer science shows the highest maximum shift in ranking (39), while the lowest maximum shift occurs in Agricultural and veterinary sciences (16). In the other areas, substantial increments are seen in Earth sciences (35) and Industrial and information engineering (32). Mathematics and computer science results as the area with least alignment between the two ranking.

| UDA | Variations | Max | Average | Median |
|---|---|---|---|---|
| Mathematics and computer science | 50 out of 51 (98.0%) | 39 | 9.22 | 8.0 |
| Physics | 40 out of 45 (88.9%) | 20 | 6.53 | 6.0 |
| Chemistry | 44 out of 45 (97.8%) | 30 | 7.44 | 5.0 |
| Earth sciences | 36 out of 39 (92.3%) | 35 | 6.67 | 5.0 |
| Biology | 47 out of 50 (94.0%) | 23 | 4.88 | 3.5 |
| Medicine | 40 out of 42 (95.2%) | 22 | 6.26 | 5.0 |
| Agricultural and veterinary sciences | 26 out of 29 (89.7%) | 16 | 5.79 | 5.0 |
| Civil engineering and architecture | 35 out of 38 (92.1%) | 28 | 7.74 | 6.0 |
| Industrial and information engineering | 42 out of 45 (93.3%) | 32 | 6.62 | 4.0 |

*Table 4: Statistics for the differences in ranking for FSS for universities between 2004-2008 and 2001-2003*

Table 5 presents, for every UDA and for each of the 63 universities analyzed, the shifts in rank expressed in quintile, for the FSS indicator. For privacy reasons, the names of the universities is not shown. Overall, the table shows that 82.5% of universities show variations in ranking by classes (quintiles) between the two time periods considered.

Examining individual universities, the greatest positive increment in rank (+4) is seen for UNIV_13 Earth sciences and UNIV_35 in Civil engineering and architecture. There are seven universities that show a positive shift of three quintiles: UNIV_4 in Civil engineering and architecture, UNIV_15 and _36 in Mathematics and computer science, UNIV_19 in Chemistry, UNIV_26 and _48 in Medicine, UNIV_54 in Industrial and information engineering. Concerning negative shifts, the greatest (-4 quintiles) occurs in three cases: UNIV_1 in Mathematics and computer science, UNIV_19 in Industrial and information engineering and UNIV_60 in Civil engineering and architecture. There are five universities that drop 3 quintiles: UNIV_2 and 16 in Earth sciences; UNIV_27 in Chemistry; UNIV_31 in Civil engineering and architecture and UNIV_36 in Agricultural and veterinary sciences.

If the shifts in quintile in all UDAs for each of the 63 universities are summed (last column, Table 5) we observe that 39.7% of the institutions show a negative balance, 42.9% a positive balance, and only 17.5% have a nil total shift. The university that shows the greatest negative total shift is UNIV_42, with a value of -7. UNIV_19 and UNIV_62 register a total shift of -5; while six institutions (UNIV_1, _10, _17, _32, _47, _58) show a total balance of -4.

The universities that show the greatest positive total shift are UNIV_4 and _35, with values of +6. Then follow UNIV_34, with +5, and UNIV_25, _36 and _54 with a total shift of +4.

If instead we consider the single UDAs (last line, Table 5), we observe that Biology has the lowest percentage of universities (38.0%) that show variations in rank (by quintile), while Agricultural and veterinary sciences shows the maximum (65.5%).

Further, the percentage of universities (of the total of universities active in each UDA) that show positive variations oscillates between the 18.0% of Biology and 34.2% of Civil engineering and architecture. In the case of negative variations, the values fall between a minimum of 20.0% for Biology and a maximum of 34.5% for Agricultural and veterinary sciences.

| University | Mathematics and computer science | Physics | Chemistry | Earth sciences | Biology | Medicine | Agricultural and veterinary sciences | Civil engineering and architecture | Industrial and information engin. | Tot. |
|---|---|---|---|---|---|---|---|---|---|---|
| Univ_1 | -4 | n.a. | n/a | n.a. | n.a. | n.a. | n.a. | 0 | 0 | -4 |
| Univ_2 | -1 | 0 | 2 | -3 | n.a. | n.a. | n.a. | 1 | -1 | -2 |
| Univ_3 | 1 | 0 | 1 | n.a. | n.a. | n.a. | n.a. | 0 | 0 | 2 |
| Univ_4 | 2 | 0 | 1 | -1 | n.a. | n.a. | n.a. | 3 | 1 | 6 |
| Univ_5 | 0 | 0 | n.a. | n.a. | n.a. | n.a. | n.a. | n.a. | n.a. | 0 |
| Univ_6 | 0 | 0 | n.a. | n.a. | 0 | n.a. | n.a. | n.a. | n.a. | 0 |
| Univ_7 | n.a. | n.a. | n.a. | n.a. | n.a. | n.a. | 0 | n.a. | 0 | 0 |
| Univ_8 | 0 | -2 | 0 | n.a. | 1 | 1 | n.a. | 0 | 0 | 0 |
| Univ_9 | n.a. | n.a. | n.a. | n.a. | n.a. | 2 | n.a. | n.a. | n.a. | 2 |
| Univ_10 | -1 | n.a. | -1 | -1 | -1 | n.a. | n.a. | n.a. | 0 | -4 |
| Univ_11 | n.a. | n.a. | n.a. | n.a. | n.a. | n.a. | n.a. | n.a. | 0 | 0 |
| Univ_12 | 0 | -1 | n.a. | n.a. | 0 | -1 | 0 | n.a. | n.a. | -2 |
| Univ_13 | n.a. | n.a. | 1 | 4 | 0 | n.a. | -2 | n.a. | n.a. | 3 |
| Univ_14 | 0 | 1 | 1 | n.a. | 0 | 0 | n.a. | n.a. | n.a. | 2 |
| Univ_15 | 3 | 0 | 0 | -1 | -1 | -1 | n.a. | 1 | 0 | 1 |
| Univ_16 | n.a. | n.a. | n.a. | -3 | 0 | n.a. | n.a. | n.a. | 1 | -2 |
| Univ_17 | -2 | 1 | -2 | 1 | 0 | n.a. | -1 | 0 | -1 | -4 |
| Univ_18 | n.a. | n.a. | 0 | n.a. | 2 | n.a. | 0 | n.a. | n.a. | 2 |
| Univ_19 | -1 | -1 | 3 | -1 | 0 | 0 | n.a. | -1 | -4 | -5 |
| Univ_20 | 1 | 0 | 0 | 0 | 0 | -1 | 2 | n.a. | 0 | 2 |
| Univ_21 | 0 | n.a. | n.a. | n.a. | n.a. | n.a. | n.a. | n.a. | 0 | 0 |
| Univ_22 | 0 | 0 | n.a. | n.a. | 0 | 1 | n.a. | -1 | 1 | 1 |
| Univ_23 | 0 | 0 | 1 | 0 | -1 | 0 | n.a. | -1 | 0 | -1 |
| Univ_24 | n.a. | n.a. | n.a. | n.a. | n.a. | n.a. | n.a. | -1 | -1 | -2 |
| Univ_25 | 2 | -1 | 0 | 0 | 0 | 1 | 2 | -1 | 1 | 4 |
| Univ_26 | n.a. | n.a. | n.a. | n.a. | 0 | 3 | 0 | n.a. | n.a. | 3 |
| Univ_27 | -1 | 1 | -3 | 0 | 0 | 0 | n.a. | 0 | 0 | -3 |

| University | Mathematics and computer science | Physics | Chemistry | Earth sciences | Biology | Medicine | Agricultural and veterinary sciences | Civil engineering and architecture | Industrial and information engin. | Tot. |
|---|---|---|---|---|---|---|---|---|---|---|
| Univ_28 | 0 | -1 | 0 | 1 | 1 | 0 | 1 | 0 | 1 | 3 |
| Univ_29 | 0 | -1 | -1 | 0 | 0 | 0 | 2 | n.a. | 0 | 0 |
| Univ_30 | 1 | -1 | 1 | 0 | 0 | 0 | n.a. | n.a. | n.a. | 1 |
| Univ_31 | 0 | 0 | 0 | 1 | 0 | 2 | 2 | -3 | 0 | 2 |
| Univ_32 | 0 | 0 | -1 | -1 | 1 | 0 | -1 | -1 | -1 | -4 |
| Univ_33 | 0 | n.a. | n.a. | -1 | 0 | n.a. | n.a. | 0 | 0 | -1 |
| Univ_34 | 0 | 1 | 2 | 1 | 0 | 0 | 1 | 0 | 0 | 5 |
| Univ_35 | -1 | 0 | 1 | 0 | 0 | 0 | 0 | 4 | 2 | 6 |
| Univ_36 | 3 | 1 | 0 | 0 | 1 | 0 | -3 | 1 | 1 | 4 |
| Univ_37 | 2 | -1 | -1 | 1 | 1 | 0 | n.a. | 2 | -1 | 3 |
| Univ_38 | 0 | 0 | 1 | 1 | -1 | 0 | -1 | 1 | -1 | 0 |
| Univ_39 | -1 | 0 | 0 | 0 | 0 | 0 | 1 | -1 | 0 | -1 |
| Univ_40 | -1 | -1 | 0 | n.a. | 0 | 0 | n.a. | 1 | 0 | -1 |
| Univ_41 | n.a. | n.a. | -1 | 0 | 0 | 0 | -1 | n.a. | n.a. | -2 |
| Univ_42 | -1 | 0 | -1 | -2 | -2 | -1 | n.a. | n.a. | 0 | -7 |
| Univ_43 | n.a. | n.a. | n.a. | n.a. | n.a. | n.a. | 2 | n.a. | n.a. | 2 |
| Univ_44 | 0 | 1 | -1 | 0 | 1 | 0 | 0 | n.a. | n.a. | 1 |
| Univ_45 | 0 | 0 | 1 | n.a. | n.a. | n.a. | n.a. | 1 | 0 | 2 |
| Univ_46 | -1 | -1 | 2 | 1 | 0 | -1 | n.a. | 0 | 0 | 0 |
| Univ_47 | 0 | 0 | -1 | 2 | -2 | -1 | -1 | -1 | 0 | -4 |
| Univ_48 | -1 | 2 | -2 | -2 | -1 | 3 | n.a. | n.a. | n.a. | -1 |
| Univ_49 | 0 | n.a. | n.a. | n.a. | 0 | 1 | 2 | n.a. | -1 | 2 |
| Univ_50 | 0 | 2 | 0 | 1 | -1 | -1 | n.a. | 2 | n.a. | 3 |
| Univ_51 | n.a. | n.a. | n.a. | n.a. | 0 | -2 | n.a. | n.a. | n.a. | -2 |
| Univ_52 | -1 | n.a. | n.a. | n.a. | n.a. | n.a. | 0 | 1 | -1 | -1 |
| Univ_53 | 0 | 0 | 0 | 0 | 0 | n.a. | n.a. | 1 | -1 | 0 |
| Univ_54 | 1 | 2 | 0 | 0 | 0 | -2 | n.a. | 0 | 3 | 4 |
| Univ_55 | -1 | 1 | 0 | n.a. | 1 | 1 | n.a. | n.a. | n.a. | 2 |
| Univ_56 | 0 | 1 | 0 | -1 | 0 | 0 | 0 | 0 | 2 | 2 |
| Univ_57 | -1 | 1 | -2 | 1 | -1 | 0 | 0 | 1 | n.a. | -1 |
| Univ_58 | -1 | 1 | -1 | 0 | 0 | 0 | -1 | 0 | -2 | -4 |
| Univ_59 | 1 | 0 | 0 | 0 | 0 | 0 | n.a. | 0 | 0 | 1 |
| Univ_60 | 0 | -1 | 1 | 1 | 1 | 0 | -1 | -4 | 1 | -2 |
| Univ_61 | 2 | 0 | -1 | n.a. | -1 | n.a. | 0 | 0 | -1 | -1 |
| Univ_62 | 0 | -1 | 0 | -1 | 0 | -1 | -2 | 0 | 0 | -5 |
| Univ_63 | n.a. | n.a. | n.a. | n.a. | 0 | 0 | n.a. | n.a. | n.a. | 0 |
| Variations | 52.9% | 55.6% | 62.2% | 61.5% | 38.0% | 45.2% | 65.5% | 60.5% | 48.9% | 82.5% |

*Table 5: Variations in rank (by quintile) for UDAs at each university, for the indicator FSS*
*"n.a." means that the University has no research staff in the UDA or that it consists of less than six scientists*

It is possible to enter into a greater depth of detail, to evaluate direction and importance of changes at the level of single UDAs. Table 6 presents the example of the Biology UDA. The table shows that, from 2001-2003 to 2004-2008, the rank of 31 of the 50 universities considered remained stable, in terms of quintile (sum of terms along the main diagonal of the matrix). Ten of the universities show variation towards lower quintiles and nine shift towards higher quintiles. Of the 10 universities that are at the top for performance ("very high") over the 2001-2003 triennium, two descend by one quintile (to "high") and one drops by two quintiles (to "medium"). At the opposite extreme, of the 10 universities that, for the first triennium, classified in the lowest quintile ("very low"), two change their status in the next period, arriving in the next quintile up ("low"). The changes in rank primarily concern shifts between adjacent quintiles, with the notable exception of one university that shifts from the bottom to the top quintile.

|  | Performance | 2004-2008 | | | | | |
|---|---|---|---|---|---|---|---|
|  |  | Very high | High | Medium | Low | Very low | Total |
| 2001-2003 | Very high | 7 | 2 | 1 | 0 | 0 | 10 |
|  | High | 2 | 6 | 2 | 0 | 0 | 10 |
|  | Medium | 0 | 2 | 5 | 3 | 0 | 10 |
|  | Low | 1 | 0 | 2 | 5 | 2 | 10 |
|  | Very low | 0 | 0 | 0 | 2 | 8 | 10 |
|  | Total | 10 | 10 | 10 | 10 | 10 | 50 |

*Table 6: Changes in performance (quintiles) of universities active in the Biology UDA, for the indicator FSS*

## 5. Analysis at the SDS level

The analyses in the previous section can be further detailed at the level of the individual SDS. This could be useful to the management of any given university to permit their understanding of the contribution of individual SDSs to the performance variation of an entire UDA. Table 7 presents the example of the variations in rank (quintiles) for FSS, for the SDSs of the Industrial and information engineering UDA at UNIV_19, where the shift in UDA ranking was -4. The table shows that seven SDSs out of 25 (28.0% of total) do not vary in rank. Of the 18 remaining SDSs, 14 show negative shifts (56.0% of total) and only 4 show positive shifts (16.0% of total). The maximum positive shift is 2, registered for ING-IND/11 and ING-IND/35; the maximum negative shift is -4, registered for ING-IND/14.

| SDS | Δ rank (quintile) | SDS | Δ rank (quintile) |
|---|---|---|---|
| ING-IND/09 | -3 | ING-IND/27 | 0 |
| ING-IND/10 | -2 | ING-IND/31 | 0 |
| ING-IND/12 | 1 | ING-IND/32 | -3 |
| ING-IND/13 | -3 | ING-IND/33 | -1 |
| ING-IND/14 | -4 | ING-IND/35 | 2 |
| ING-IND/15 | -2 | ING-INF/01 | -1 |
| ING-IND/16 | -2 | ING-INF/02 | -1 |
| ING-IND/17 | 0 | ING-INF/03 | 1 |
| ING-IND/22 | 0 | ING-INF/04 | -1 |
| ING-IND/24 | 0 | ING-INF/05 | -1 |
| ING-IND/25 | -2 | ING-INF/07 | -3 |
| ING-IND/26 | 0 | | |

*Table 7: Variations in rank (quintiles) for FSS, for SDSs of the Industrial and information engineering UDA of an Italian university (UNIV_19)*

The evaluation system developed by the authors permits arrival at a still deeper level of detail, with the unit of observation being the individual scientist. It is thus possible to understand who contributes, and how much, to the performance variation of a given SDS over consecutive time periods.

**6. Comparison of performance shifts per indicator**

The analyses presented above refer to the FSS indicator, which synthesizes all the relevant dimensions of performance (quantity, quality, contribution), but they can easily be repeated for the other indicators. This permits detection of any potential shifts in focus from "quantity" to "quality", or vice versa. Table 8 returns to the example of SDSs of the Industrial and information engineering UDA of an Italian university, showing the shifts in rank (by quintile) registered in correspondence to P, FP and AQ.

In general, the table shows that a loss of rank for P (13 cases) never corresponds to a gain in positions for AQ, with the exception of ING-IND/17 and ING-INF/02.

There are 7 SDSs where there are losses in rank for all three indicators: ING-IND/10, /13, /15, /16, /25, /32 and ING-INF/07. However, there is only one SDS that shows gains in rank for all three indicators: ING-IND/35.

| SDS | P | FP | AQ |
|---|---|---|---|
| ING-IND/09 | -2 | -2 | 0 |
| ING-IND/10 | -2 | -1 | -2 |
| ING-IND/12 | 0 | 0 | 0 |
| ING-IND/13 | -2 | -2 | -1 |
| ING-IND/14 | -2 | -4 | 0 |
| ING-IND/15 | -2 | -2 | -1 |
| ING-IND/16 | -2 | -3 | -1 |
| ING-IND/17 | -1 | 0 | 2 |
| ING-IND/22 | -1 | 0 | -1 |
| ING-IND/24 | 0 | 0 | 3 |
| ING-IND/25 | -1 | -1 | -3 |
| ING-IND/26 | 0 | 0 | -3 |
| ING-IND/27 | -1 | 0 | 0 |
| ING-IND/31 | 0 | 0 | -1 |
| ING-IND/32 | -3 | -3 | -2 |
| ING-IND/33 | 0 | 0 | 0 |
| ING-IND/35 | 2 | 3 | 1 |
| ING-INF/01 | 0 | -1 | -1 |
| ING-INF/02 | -3 | -2 | 3 |
| ING-INF/03 | 1 | -1 | 1 |
| ING-INF/04 | 0 | -1 | 1 |
| ING-INF/05 | -1 | -2 | 0 |
| ING-INF/07 | -1 | -1 | -2 |

*Table 8: Variation in rank (quintiles) for SDSs in the Industrial and information engineering UDA of an Italian university (UNIV_19), for various performance indicators*

Finally, for three SDSs (ING-INF/02, /03 and /04), there is a positive shift for AQ together with a negative shift for FP. This phenomenon could be explained in the light of the results presented in the literature (Franceschet and Costantini, 2010; Abramo et al., 2009; He et al., 2009), which show a positive correlation between intensity of collaboration (with parity in P this indicates a lower FP) and impact of the scientific output. In the case of these specific SDSs, the research staff have, through greater collaboration, improved the average quality of their output.

## 7. Conclusions

In order to address global competitiveness, there is currently a world-wide tendency to orient science policy ever more towards performance and competition mechanisms, with the objective of fostering higher efficiency in research institutions. International experience has shown that performance-based evaluation conducted at periodic intervals indeed improves performance of these systems. The critical element present in studies that evaluate the impact of these assessment exercises resides in the difficulty of fully evaluating the extent of the changes that occur in each organization between two different evaluation exercises, even when these are consecutive. On the one hand there are often changes in the rules of the game, on the part of the policy maker, in the attempt to improve the incentive system through placing more or less emphasis on various aspects of scientific performance (quantity or quality), or through differing representativity of the share of output evaluated, etc. On the other hand there is the difficulty of separating the effects flowing from policy interventions concerning assessment exercises from those deriving from endogenous factors, such as natural performance variation over time in the research institutions.

The objective of this work has been to measure the natural variation in research institution performance under two important conditions characteristic of the Italian case: i) the use of a methodology and evaluative criteria that remained invariable over time; ii) the absence of policy interventions sufficient to affect the work of scientists over the two periods examined.

The work examined research performance in the hard sciences at Italian universities in two consecutive periods: 2001-2003 and 2004-2008. The results show that, between the first and second period, both the productivity and the average impact registered at the level of individual scientist increased, in all the disciplinary areas considered. At the level of university it was possible to observe significant variations in rank for all the bibliometric indicators considered and in all disciplinary areas. The proposed system also permits reaching a deeper level of detail: the analysis conducted at the level of disciplinary sectors permits understanding which of those belonging to a given area have determined, and in what measure, the variations observed at the higher level. The comparative analysis for pairs of indicators permits further identification of shifts of focus from "quantity" to "quality", or vice versa.

In the absence of policy interventions that could have altered the status of the system, it was possible to thus quantify the extent of the natural variation in performance of the research organizations. This makes it possible to successively evaluate the "net effect" of policy interventions eliminating the endogenous component, in order to quantify their real impact.

At this point it becomes more manageable to tackle the other critical problem of evaluation mentioned in the introduction, meaning the analysis of results from a research assessment in the presence of distortions introduced by variations in criteria, indicators, and/or methodology employed in the evaluation. The authors plan to work on exactly this analysis, in the near future.